\documentclass[conference]{IEEEtran}
\IEEEoverridecommandlockouts
\pdfoutput=1
\usepackage{cite}

\ifCLASSINFOpdf
  \usepackage[pdftex]{graphicx}
\else
\fi
\usepackage{algorithm}
\usepackage[noend]{algpseudocode}
\usepackage[tight,footnotesize]{subfigure}

\usepackage[font=footnotesize]{subfig}
\usepackage{stfloats}

\usepackage{url}

\def\BibTeX{{\rm B\kern-.05em{\sc i\kern-.025em b}\kern-.08em
    T\kern-.1667em\lower.7ex\hbox{E}\kern-.125emX}}

\usepackage{multirow}

\usepackage{multirow}
\usepackage{tikz}
\usepackage{lipsum}

\newcommand\copyrighttext{%
  \footnotesize \textcopyright {2020 IEEE}
  DOI: \url{https://www.doi.org/10.1109/IRI49571.2020.00021}}
\newcommand\copyrightnotice{%
\begin{tikzpicture}[remember picture,overlay]
\node[anchor=south,yshift=10pt] at (current page.south) {\fbox{\parbox{\dimexpr\textwidth-\fboxsep-\fboxrule\relax}{\copyrighttext}}};
\end{tikzpicture}%
}

\begin{document}
%
\title{Distributed Differentially Private Mutual Information Ranking and Its Applications}


\author{\IEEEauthorblockN{Ankit Srivastava, Samira Pouyanfar, Joshua Allen, Ken Johnston, Qida Ma}
\IEEEauthorblockA{Azure COSINE Core Data\\
Microsoft\\
Redmond, United States\\
Emails: \{ankitsri,sapouyan,joshuaa,kenj,t-qidma\}@microsoft.com}
}


%


\maketitle
\copyrightnotice

\begin{abstract}
Computation of Mutual Information (MI) helps understand the amount of information shared between a pair of random variables. Automated feature selection techniques based on MI ranking are regularly used to extract information from sensitive datasets exceeding petabytes in size, over millions of features and classes. Series of one-vs-all MI computations can be cascaded to produce n-fold MI results, rapidly pinpointing informative relationships. This ability to quickly pinpoint the most informative relationships from datasets of billions of users creates privacy concerns. In this paper, we present Distributed Differentially Private Mutual Information (DDP-MI), a privacy-safe fast batch MI, across various scenarios such as feature selection, segmentation, ranking, and query expansion. This distributed implementation is protected with global model differential privacy to provide strong assurances against a wide range of privacy attacks. We also show that our DDP-MI can substantially improve the efficiency of MI calculations compared to standard implementations on a large-scale public dataset.
\end{abstract}


%
\IEEEpeerreviewmaketitle

\section{Introduction}
Nowadays, server-side logs of opted-in user interactions and feedback are collected, stored, and analyzed by numerous organizations. These logs are queried and used for a variety of purposes, such as improving the organizations’ products and services, making them more secure, and gaining business intelligence via metrics and dashboards. Measures such as Mutual Information (MI) \cite{cover2012elements} can estimate the amount of information shared between features and dimensions stored in these logs. Examples of such applications of MI ranking include finding distinguishing websites (features) visited per country (partitions), distinguishing feedback topics received per device make or model, distinguishing products purchased across regions, and distinguishing movies watched in a zip code. 

Mutual Information of two discrete random variables X and Y is calculated as per Eq. (\ref{mutualinformation}).
\begin{equation}\label{mutualinformation}
I(X;Y) = \sum_{y \in \mathcal Y} \sum_{x \in \mathcal X} { p_{(X,Y)}(x,y) \log{ \left(\frac{p_{(X,Y)}(x,y)}{p_X(x)\,p_Y(y)} \right) }}    
\end{equation}

where $p_{(X,Y)}$ is the joint probability mass function of $X$ and $Y$, and $p_X$ and $p_Y$ are the marginal probability mass functions of $X$ and $Y$. MI is a common measure of association used in machine learning applications, with one desirable property being that it is scale invariant, reducing the need for feature normalization. 

There is a need to compute such measures on batch datasets whose size can scale to petabytes. Such large datasets are stored in distributed computing environments such as Hadoop \cite{friedman2009sql} and Spark \cite{tejada2017mastering}. Languages such as Scala \cite{odersky2004overview}, PySpark \cite{spark2018apache} are suitable for querying large volumes of data stored in distributed computing environments in a reasonable time. However, computing the mutual information on exact aggregates is vulnerable to re-identification attacks, especially with the growing availability of auxiliary information about individuals \cite{liu2015set}. Therefore, there is a need to compute MI such that it preserves individual privacy while having minimal impact on the accuracy of the ranked features per class ordered by their information gain. 

Differential Privacy (DP) \cite{dwork2006calibrating} is the gold standard in preserving end-user privacy.
An algorithm $\mathcal{A}$ provides $\epsilon$-differential privacy if for all neighboring datasets $D_1$ and $D_2$ that differ on a single element (i.e. the data of one person) and for all measurable subsets $Y$ of $\mathcal{Y}$, the condition in Eq. (\ref{diffpriv}) is met. 
\begin{equation}\label{diffpriv}
\Pr[\mathcal{A}(D_1) \in Y] \leq e^\epsilon \cdot \Pr[\mathcal{A}(D_2) \in Y],
\end{equation}

This is a standard definition of differential privacy, indicating that the output of the mutual information ranking for datasets that differ by a single individual will be practically indistinguishable, bounded by the privacy parameter epsilon.

This paper introduces the application of Distributed Differentially Private Mutual Information (DDP-MI) ranking and its use across a variety of case studies. The novel contribution of this paper is to compute information gain using MapReduce with billions of instances reported across millions of features and classes in a distributed computing environment such that it provides strong assurances against privacy attacks. In addition, we demonstrate the use of cascaded MI ranking executions one after the other. Our implementation can run on standard cloud data processing infrastructure, and may efficiently be deployed in environments with secure, trusted computing architectures \cite{NIPS2019_9517}.  Our methodology can scale and, at the same time, protects end-user privacy. We present privacy-compliant case studies on how information gain criterion can be used for various business intelligence scenarios.

\section{Related Work}
\subsection{Scalable Mutual Information Computation}
MI is a well-studied criterion for automated feature selection \cite{peng2005feature} \cite{vergara2014review}. There has been an existing implementation of MI Measure to rank inferred literature relationships using traditional SQL and Visual Basic technology stack \cite{wren2004extending}. Shams and Barnes provided a method to speed up MI computation for image datasets on GPUs by approximating probability mass functions \cite{shams2007speeding}. The MapReduce batch computation of MI gain was demonstrated at cloud scale by Zdravevski et al. \cite{zdravevski2015feature} \cite{zdravevski2015parallel}. Li et al. \cite{li2019high} provided a high throughput computation of Shannon MI in a multi-core setting. 

\subsection{Differential Privacy}
DP \cite{dwork2006calibrating} has been an active area of research for more than a decade, gaining importance with increased data collection and privacy regulations such as GDPR \cite{voigt2017eu} and CCPA \cite{de2018guide}, HIPAA \cite{assistance2003summary} and FERPA \cite{curtis1974family}.  Differential Privacy has emerged as the gold standard definition of privacy, and provides formal assurances against a wide range of reconstruction and re-identification attacks, even when adversaries have auxiliary information.

U.S. Census Bureau announced, via its Scientific Advisory Committee, that it would protect the publications of the 2018 End-to-End Census Test using differential privacy \cite{abowd2018us}. Tools such as the Open Differential Privacy platform \cite{king2019harvard} aim to ease deployment of differential privacy for common scenarios.  There has been recent research that enables global differential privacy on SQL analytic workloads \cite{kotsogiannis2019architecting} \cite{kotsogiannis2019privatesql} \cite{wilson2019differentially}.  Tools that are agnostic to compute targets and can scale to petabyte scale distributed computing environments enable the use of differentially private aggregates and censoring of rare dimension for computation of mutual information ranking.

To the best of our knowledge, this is the first work on distributed MI ranking in the setting of aggregated global model DP. 

\section{Methodology}
The methodology of computing distributed differentially private MI ranking of features across partitions is illustrated in this section. It starts with collecting observations about entities such as users or devices across a set of features they interact with, such as movies watched or products purchased. The observations can be bucketed across partitions using user grain attributes such as zip code, country, device model, or time span of observation. The observation can be any discrete numeric measure such as dwell time or visit count.

\subsection{Input Data}
Table~\ref{table:1} illustrates the schema to compute MI for a dataset of user records across features and partitions. The data can be stored in any environment that supports SQL-92 grammar \cite{allen2020whitenoise} \cite{opendp2020harvardmicrosoft}\cite{opendp2020whitenoisesystem} to perform private aggregations on it. 

\begin{table}[]
\centering
\begin{tabular}{|l|l|l|l|}
	\hline
	\textbf{Id} & \textbf{Feature} & \textbf{Partition Label} & \textbf{Observation}\\
	\hline
    User 1 & Feature 1 & Partition 1 & 200.0\\
	\hline
    User 2 & Feature 1 & Partition 2 & 100.0\\
	\hline
    User 3 & Feature 2 & Partition 3 & 50.0\\
	\hline
    User 3 & Feature 3 & Partition 3 & 270.0\\
    \hline
    ... & ... & ... & ... \\
    \hline
    User N & Feature F & Partition P & 150.0\\
	\hline
\end{tabular}
\caption{Input Data Schema for distributed MI}
\label{table:1}
\end{table}

\subsection{Differentially Private SQL for Marginal and Joint Probabilities}
To compute the mutual information, we need to compute aggregates for the marginal probability of each feature and category, as well as the joint probabilities.  These are simply sums, which are supported by SUM and GROUP BY in the SQL language. Publicly-available implementations of differentially private SQL have support for releasing these sums in a privacy-preserving manner.

When applying differential privacy, noise is calibrated based on the sensitivity and maximum number of records each user can contribute.  Publicly-available implementations of differential privacy will automatically clamp values to the specified sensitivity range, sample values, and calibrate noise to the given sensitivity and contribution limit.  

When computing the marginal and joint probabilities, special care must be taken to avoid leaking infrequent dimensions.  Because the features typically come from an unbounded domain, rare features may become uniquely identified with individuals.  Some combinations of feature and category may particularly uniquely identifying.  Publicly available implementations of differential privacy operate by dropping infrequent dimensions that might violate privacy. These dropped dimensions are typically not relevant to mutual information ranking.

Because joint and marginal probabilities are computed at different levels of aggregation, it is common for joint members of a larger marginal to be dropped by differentially private processing.  This means that marginal probabilities computed by summing over the joint probabilities will be inaccurate.  One easy way to fix this is to compute the marginals separately from the joint probabilities, and pay a separate epsilon cost for each query, being sure to spread the privacy budget appropriately across all queries.  Another approach is to aggregate all dropped dimensions to an  ``other'' category.

Publicly-available implementations of differentially private SQL support censoring of rare dimensions by thresholding noisy counts\cite{korolova2009releasing}.  For settings where each user can be associated with a large number of different features or categories, Differentially Private Set Union can be used in a preprocessing step \cite{gopi2020differentially}.

In a map-reduce setting, the computation of all joint probabilities is a single pass of cost O(n log n).  Because of the dimension censoring issue mentioned earlier, the marginals may be computed separately at similar cost.  Clamping, adding noise, and dropping dimensions are constant cost, while reservoir sampling may be easily implemented in a preprocessing step of O(n log n).  Some underlying engines may enforce the contribution and sensitivity properties through database integrity constraints, in which case these checks can be skipped.

\subsection{MI Ranking}

Once the DP aggregates are available per feature and partition, we pass them through the distributed implementation of MI ranking mentioned in Algorithm~\ref{alg:miranking}. Binary MI and Single MI algorithms are called within Algorithm~\ref{alg:miranking} as helper utility methods that perform the actual computation of mutual information gain for every feature, partition combination.

\begin{algorithm}[]
\caption{MI Ranking algorithm}\label{alg:miranking}
\begin{algorithmic}[1]
\Procedure{MI Ranking}{Feature, Partition, Observation}
\State $obv_{normalized} \gets obv_{private}$ GROUP BY per Feature, Partition
\State $p_{xy} \gets obv_{normalized}$ $>$ 0
\State $p_x \gets obv_{private}$ GROUP BY per Feature
\State $p_y \gets obv_{private}$ GROUP BY per Partition
\State $mi \gets$ CalcMI($p_x, p_y, p_{xy})$ GROUP BY per Feature, Partition
\State $ordered\_mi \gets$ Order by $mi$ in descending order
\State $\Delta \gets (p_x - p_{xy})$
\State $x \gets p_{xy} / p_y$
\State $y \gets \Delta / (1 - p_y)$
\State $direction \gets x > y$ ? ``Presence'' : ``Absence''\\
\textbf{return} $ordered\_mi$, $direction$ per Feature, Partition
\EndProcedure
\end{algorithmic}
\end{algorithm}

We first take the differentially private sum of observations for each feature and partition combination. This sum is normalized into a joint probability by dividing the summed observations for each feature and partition combination by the total observations across all features and partitions.  Only observations greater than zero are considered for computation of joint probabilities $p_{xy}$. Similarly, we get differentially private marginal probabilities $p_{x}$ and $p_{y}$. Grouping by partitions takes care of one-vs-all setting of MI computation. Using computation of direction, we can find the presence or absence of a feature that was highly distinguishing of a partition. For example, the presence of commercial apps and the absence of consumer apps being used on a device can indicate the device belongs to the commercial segment. 

\begin{algorithm}[]
\caption{Binary MI Algorithm}\label{alg:binarymi}
\begin{algorithmic}[1]
\Procedure{CalcMI}{$p_x, p_y, p_{xy}$}
\State $p_{\neg x} \gets 1.0 - p_x$ 
\State $p_{\neg y} \gets 1.0 - p_y$ 
\State $p_{x\neg y} \gets p_x - p_{xy}$ 
\State $p_{y\neg x} \gets p_y - p_{xy}$
\State $p_{\neg{xy}} \gets 1.0 - p_{x} - p_{y}+ p_{xy} $ \\
\textbf{return}
\Statex CalcSingleMI$(p_x, p_y, p_{xy})$ +
CalcSingleMI$(p_x, p_{\neg y}, p_{x\neg y})$ +
CalcSingleMI$(p_{\neg x}, p_y, p_{y\neg x})$ +
CalcSingleMI$(p_{\neg x}, p_{\neg y}, p_{\neg{xy}})$
\EndProcedure
\end{algorithmic}
\end{algorithm}

For cases where a feature is in one class but not the others, some arms of the Singe MI computation return zero.  If the joint count is non-zero, but low, and the marginal count is very large, the probability computation can approach zero, outside the ability of standard floating point implementations to represent.  For this case, we choose a threshold tolerance parameter $tol$ (e.g. 10E-16) which probabilities do not fall below.  Note that infrequent joint probabilities with privacy exposure have been filtered out prior to this step, so remaining probabilities that fall below $tol$ will have reduced accuracy, but no additional privacy cost.

\begin{algorithm}[]
\caption{Single MI Algorithm}\label{alg:singlemi}
\begin{algorithmic}[1]
\Procedure{CalcSingleMI}{$p_x, p_y, p_{xy}, tol$}\
\If {$p_{xy} < tol$}
    \textbf{return} 0.0
\EndIf
\State $\phi \gets p_x * p_y$
\If {$\phi < tol$}
    \State $\phi \gets tol$
\EndIf\\
\textbf{return} $p_{xy} * \log(p_{xy} / \phi)$
\EndProcedure
\end{algorithmic}
\end{algorithm}

\section{Applications of MI Ranking}
In this section, several important real-world applications of DDP-MI are presented. 
\subsection{Segmentation Without Labels}
One of the early applications we applied MI is to find which Win32 applications are used predominantly by teachers and students but not by the rest of the population. We wanted to make sure such Win32 applications are available in the Microsoft Store. Since there is no easy way to get labels for teacher and student devices in Windows 10 K-12 US install base, we applied MI to find distinguishing applications for each segment. 

From domain knowledge, we knew Smart Ink is a popular Win32 application used for digitizing the markerboard and is heavily used by teacher devices, not student devices. Using that as a seed application, we identified a cohort of US K-12 devices that had opted in for Full diagnostic data and Tailored experiences.  We then categorized devices using ``Smart Ink'' in one category and the rest of K-12 US devices population in the second category. Using binary MI ranking, we identify other such applications that are distinguishing of teacher vs. student/admin devices as listed in Table~\ref{table:2}.
This process helped us to expand the cohort from just initial seed app we began with to then identify hundreds of applications that have high information gain in distinguishing teachers’ and students’ devices. This lets us build a segment of teacher and student devices. We also gained insight about student devices’ browser apps being locked down by applications like ``lockdownbrowser.exe'' for purposes of student testing. Instead of MI ranking, if we only use top apps, the browser apps appear on the top. However, they do not have high information gain between teacher and student/admin devices.

\begin{table}[]
\centering
\begin{tabular}{|l|l|}
	\hline
	\textbf{Teacher apps} & \textbf{Student / Admin apps}\\
	\hline
    notebook.exe & lockdownbrowser.exe\\
	\hline
    smartintdocumentviewer.exe & nk\_station.exe\\
	\hline
    imagemate.exe & inventor.exe\\
	\hline
    gqweb.exe & photoshop.exe\\
    \hline
    floatingtools.exe & student.exe\\
    \hline
    smartboardtools.exe & onesource.exe\\
	\hline
	grpwise.exe & sldworks.exe\\
	\hline
\end{tabular}
\caption{Top Teacher vs Student Apps}
\label{table:2}
\end{table}

\subsection{Multi-class classification at scale}
For browsers, website compatibility is a big focus area. We used sampled anonymized webpage visit logs per country from devices opted into collecting diagnostic data. There are a lot of top websites that get visited by users across countries. To understand country-specific compatibility needs, we applied MI ranking to understand websites predominantly visited by users of a specific country but not by others. MI ranking in a single run is able to identify top distinguishing websites for each of the 195 countries. We pick three arbitrary countries from the per-country results and showcase top distinguishing websites for them in Table~\ref{table:3}. 

\begin{table}[]
\centering
\begin{tabular}{|l|l|l|}
	\hline
    \textbf{US} & \textbf{India} & \textbf{China}\\
	\hline
    msn.com/market=US  & irctc.co.in & baidu.com\\
	\hline
    aol.com & google.co.in & cn.bing.com\\
	\hline
    xfinity.com & msn.com/en-in & hao.qq.com\\
	\hline
    att.com & jionetportal.co.in & bilibili.com\\
    \hline
    roblox.com & hdfcbank.com & taobao.com\\
	\hline
\end{tabular}
\caption{Top 5 Websites per Country}
\label{table:3}
\end{table}

\subsection{Query expansion}
One of the key functions of marketing is to buy keywords on search engines.  Often this is done with some basic analysis of query volumes and click-through rates on competitors adds that creates a bidding war over a small set of fairly generalized keywords.  MI allows us to identify even more distinguishing keywords, and usually, those keywords will have a lower bid threshold Cost per Click (CPC). In particular, we try to understand IoT enthusiasts who query on Bing search engine. For this purpose, we started with a seed query of ``azure iot'' and applied MI ranking to find queries distinguishing of visitors who search for ``azure iot'' in a time span of 28 days.
From just a seed query with thousands of visitors, in just hours of processing and a few days of analysis we are able to derive a breadth of related search queries that are distinguishing of IoT enthusiasts (as shown in Table~\ref{table:5}) and from there build a target audience that is suitable for ads around our service offering.

\begin{table}[]
\centering
\begin{tabular}{|l|l|}
	\hline
	\begin{tabular}[l]{@{}l@{}} \textbf{Query} \textbf{Category} \end{tabular} & \textbf{Sample Queries}\\
	\hline
    \begin{tabular}[l]{@{}l@{}} Related Azure\\ Services \end{tabular} & \begin{tabular}[l]{@{}l@{}}azure event hub, azure blob storage, windows\\ 10 iot core, azure cognitive search, azure\\ monitoring, azure pricing calculator\end{tabular}\\
    \hline
    \begin{tabular}[l]{@{}l@{}} Datasheets and\\  IoT Boards\end{tabular} & \begin{tabular}[l]{@{}l@{}} l293d (Texas Instruments), intel nuc, 74hc595\\ (Texas Instruments)\end{tabular}\\
    \hline
    \begin{tabular}{@{}l@{}}Azure IoT \\Feature Searches\end{tabular} & \begin{tabular}{@{}l@{}}iotedge, iot hub, azure sphere, azure edge\\ raspberry pi iot core, iot analytics platform\end{tabular}\\
    \hline
    \begin{tabular}[l]{@{}l@{}} Developer \\Searches \end{tabular} & \begin{tabular}{@{}l@{}}docfx (Publishing API), npm azure-iothub,\\ azure iothub sdks, azure iot dev kit, azure iot\\ c sdk docs\end{tabular}\\
    \hline
    \begin{tabular}{@{}l@{}}IoT Technology\\Standards\\ Protocols\end{tabular} & \begin{tabular}{@{}l@{}}nb-iot (Narrowband IoT), lte-cat-m1 (Low\\ Power wide-area air interface that lets connect\\ IoT and M2M devices), mqtt (lightweight\\ messaging protocol for small sensors)\end{tabular}\\
	\hline
\end{tabular}
\caption{Top Queries of Azure IoT Enthusiasts}
\label{table:5}
\end{table}

\subsection{Symmetric mutual information}
Mutual Information gain between two random variables is symmetric i.e. $I(X;Y) = I(Y;X)$. So, we can reverse MI results from distinguishing features per partition to distinguishing partitions per feature. This property of MI is interesting where we want to find specific segments where a given feature, let's say an application is used as a distinguishing application.

We can see in Table~\ref{tab:apps} that many entities are only distinguishing of a single segment (built for that segment) like Minecraft for Education is used as a distinguishing app in the K-12 segment. Some entities are distinguishing more than one segment. The feature or application might be used across all segments, but some leverage it significantly more than others. For example, given the collaborative nature of industries like Partner Professional Services, Media, Telecommunications, and Technology, applications such as Microsoft Teams are distinguishing within those segments. MI ranking works best when at least one of the sides of comparison is not too high-dimensional.  For example, it is fine to have billions of web sites compared with thousands of segments, however, comparing billions of entities on one side with billions of partitions on the other side will not perform well.

\begin{table}[]
\centering
\begin{tabular}{|l|l|}
\hline
\textbf{Label Name}               & \textbf{Applications} \\ \hline
K-12    & microsoft.minecrafteducationedition                        \\ \hline
Partner Professional Services         & teams.exe                        \\ \hline
Media \& Telcom          & teams.exe                       \\ \hline
Technology          & teams.exe                       \\ \hline
\end{tabular}
\caption{Top Partitions (Industry) Per Feature}
\label{tab:apps}
\end{table}

\subsection{N-Fold mutual information ranking}
In certain applications, looking at one entity type is not enough. For example, to build a propensity model on which organizations are likely to purchase a service, there is a need to tap into various customer touchpoints like public news articles, marketing documentation visits, service subscriptions. Thus, we propose a methodology to apply 2-fold MI on data.

\begin{figure}[h]
\includegraphics[width=0.45\textwidth]{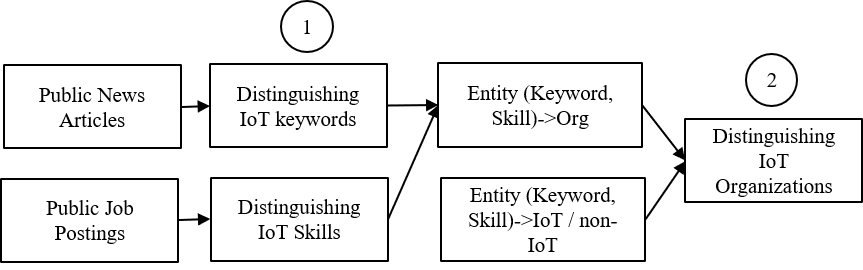}
\caption{Distinguishing IoT Companies via 2-fold MI}
\label{fig:2foldMI}
\end{figure}

Figure~\ref{fig:2foldMI} demonstrates this approach leveraged in identifying companies working in IoT space. The first set of MI runs find keywords in publicly available sources of data like news articles and job postings distinguishing of IoT. They expand from a small seed list of IoT keywords and skills. The second step is to link these identified IoT entities to organizations and run MI to identify organizations distinguishing of IoT entities. These steps can be alternated repeatedly to expand the number of associated entities discovered. Each step incurs an additional privacy cost, so epsilon must be composed across multiple folds.

\begin{table}[h]
\centering
\begin{tabular}{|l|l|}
\hline
\textbf{Industry}               & \begin{tabular}[l]{@{}l@{}}\textbf{Top MI ranked IoT}\\ \textbf{Distinguishing Organizations}\end{tabular}  \\ \hline
Manufacturing    & \begin{tabular}[l]{@{}l@{}} Intel (IoT platform),\\ Flextronics (IoT platform – SmartNexus)  \end{tabular}                       \\ \hline
\begin{tabular}[l]{@{}l@{}}Professional\\ Services \end{tabular}         & \begin{tabular}[l]{@{}l@{}}Cadillac Fairview (Commercial Real Estate\\ Leasing – IoT in Smart Buildings),\\ Altran (Industrial IoT) \end{tabular}                       \\ \hline
\begin{tabular}[l]{@{}l@{}} Higher\\ Education  \end{tabular}          & \begin{tabular}[l]{@{}l@{}} Arizona State University (invests in IoT\\with Intel – Smart Stadium, Smart Campus),\\ Yonsei University (IoT specialization on Coursera)  \end{tabular}                      \\ \hline
Retailers          & \begin{tabular}[l]{@{}l@{}} Starbucks (Hyper connected coffee shops),\\ Assurant Group (Connected Living Division)  \end{tabular}                      \\ \hline
\end{tabular}
\caption{Distinguishing IoT Companies Using 2-Fold MI Across Various Organizations}
\label{tab:2foldMI}
\end{table}

\section{Experimental Analysis}
\subsection{Differential Privacy Analysis}
To understand the effect of applying differential privacy on mutual information ranking, we use a large-scale dataset including 5,880,165 features and 22 partition labels across 382,762,990 randomly generated private user identifiers. Features and partitions correspond to a simple scenario where the maximum contribution of a private user identifier is set to one feature value and one partition label. The dataset is aggregated to get a user count grouped by features and partition labels. The aggregated dataset is passed through OpenDP WhiteNoise-System to get differentially private user count for a range of epsilon values (privacy loss parameter) from [0.1, 0.5, 1.0, 2.0, 4.0, 8.0]. The lower the epsilon value, the higher the guarantee of privacy but lower the accuracy. The six DP aggregated datasets corresponding to different epsilon values are passed through the DDP-MI algorithm. We filter to the top 10K feature and partition label combinations with the highest MI gain and compare their MI ranks with the ones they had without the use of differential privacy.

To compare the rankings, we use \{10, 25, 50, 75, 90\} percentiles of absolute difference in ranks as the evaluation metric. 50th percentile corresponds to median absolute error as per Eq. (\ref{medianabsoluteerror}) where $y$ corresponds to original MI rankings without the use of DP and $\hat{y}$ corresponds to the rankings from DDP-MI algorithm. 

\begin{equation}\label{medianabsoluteerror}
MedAE(y, \hat{y}) = median({\mid y_1 - \hat{y}_1 \mid,...,\mid y_n - \hat{y}_n \mid})  
\end{equation}

\begin{figure}[h]
    \includegraphics[width=0.5\textwidth]{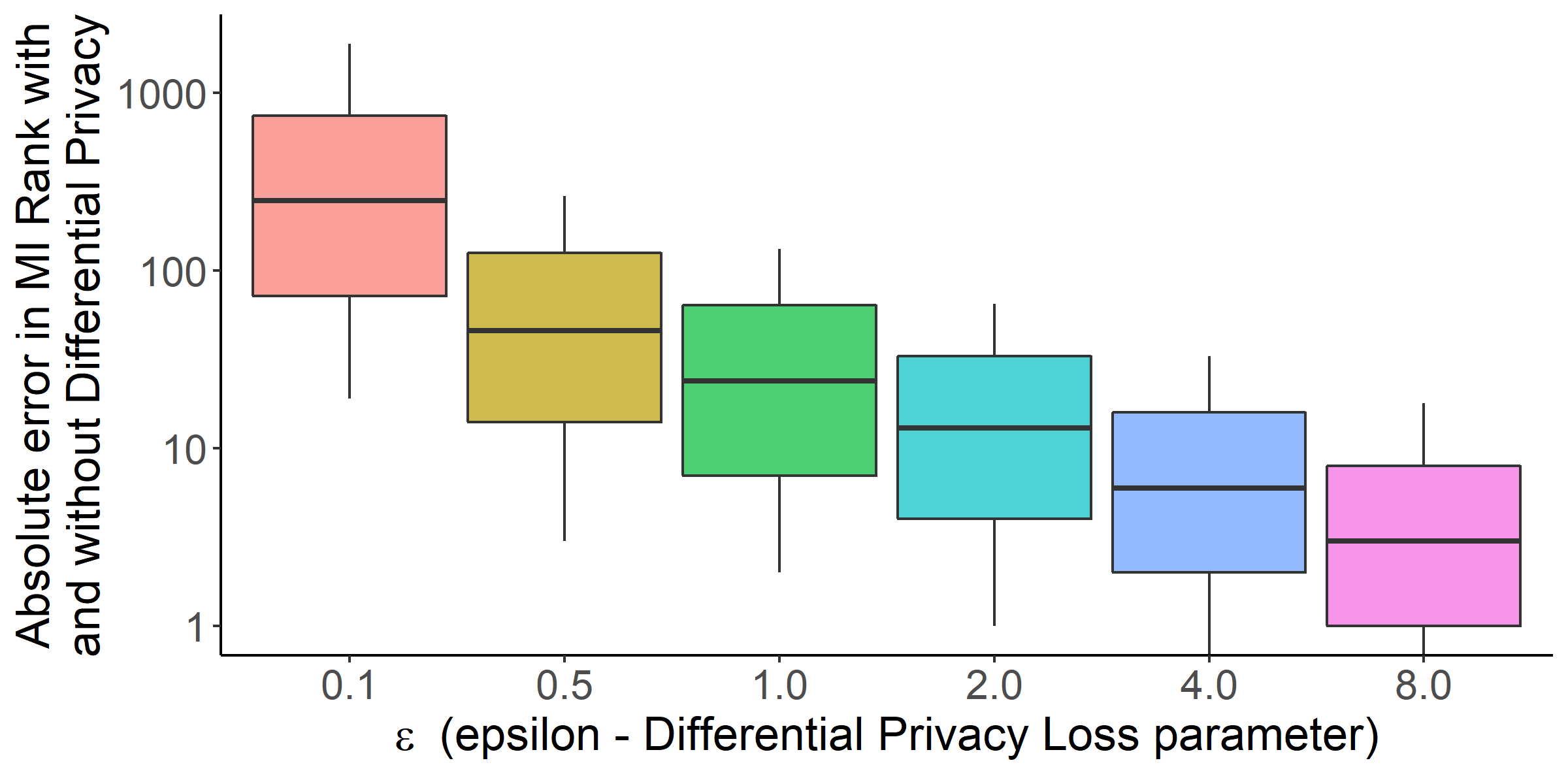}
   \caption{Comparing top 10K MI Ranks with and without DP}
   \label{fig:MIDPComparison}
\end{figure}

Figure~\ref{fig:MIDPComparison} presents the results of this comparison. As can be seen from the figure, small values of epsilon lead to relatively high absolute errors in MI ranks. Therefore, setting epsilon ($\epsilon$) to 1.0 or 2.0 is a reasonable choice for low privacy loss and low absolute error.

\begin{figure}[h]
    \includegraphics[width=0.4\textwidth]{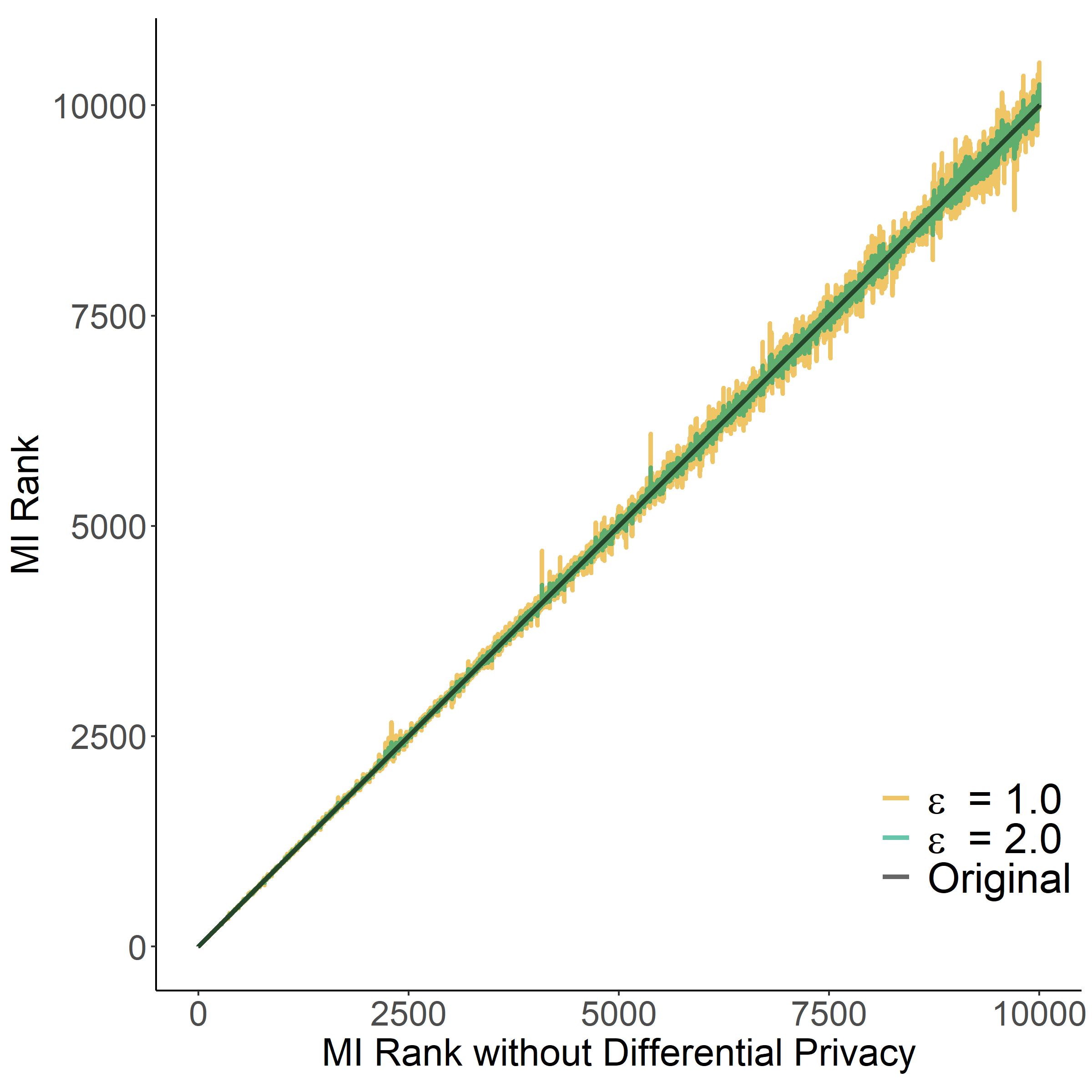}
   \caption{Stability of MI Ranks with and without DP}
   \label{fig:MIRankStability}
\end{figure}

Figure~\ref{fig:MIRankStability} further shows stability of MI ranks among top features with highest mutual information gain and the tail ranks. We can see most of the absolute error is contributed by the tail features which have low information gain. Use of differential privacy hardly has any impact on the top ranking features. The lower ranked features can potentially be pruned out. 

\subsection{Runtime Performance Analysis}
In addition to the private applications, the DDP-MI is applied on a public dataset and its runtime performance is compared with two different MI implementations on Python and R. 
This experiment show the efficiency of the DDP-MI on large-scale datasets compared to some other available work in this area. For this purpose, an academic citation network dataset~\cite{tang2008arnetminer}, collected by Arnetminer~\footnote{\url{https://aminer.org/citation}}, was utilized for this study.
This data are collected from different resources such as DBLP, ACM, and Microsoft Academic Graph (MAG). It includes major computer science journals, conference proceedings and arXiv pre-prints. 
We used Citation-network V10 from this dataset that contains 3,079,007 papers (nodes) and 25,166,994 citation relationships (edges). The information of this dataset is stored in a JSON file, and each line contains the following fields: paper id, title, authors, venue, year, number of citations, references, and abstract. In particular, we used paper id as Id, extracted words from abstract as Features, and frequency of these keywords as Observation. We also used the year of publication as Partition (category label). Using MI, we can find the most distinguishing keywords for each year.

We evaluated the runtime performance of our proposed DDP-MI with the following benchmarks: 
1) Python's ``mutual\_info\_classif'' function~\footnote{\url{https://scikit-learn.org/stable/modules/generated/sklearn.feature_selection.mutual_info_classif.html}} from sklearn package, and 2) R's ``mutinformation'' function~\footnote{\url{https://www.rdocumentation.org/packages/infotheo/versions/1.2.0/topics/mutinformation}} from infotheo library.

In the first experiment, we compared the performance of our Binary MI with DDP-MI 
on different sets of Academic Citation data (from about 3 million samples to 60 million samples). 
For binary MI, we run the MI algorithm for each partition (year) separately using one-against-all technique, while DDP-MI can automatically handle multi-partitions data. The performance results are shown in Table ~\ref{tab:performance1}. From this table, it is obvious that DDP-MI can greatly enhance the running performance compared to binary MI (about 600\%-700\%).

\begin{table}[]
\centering
\begin{tabular}{|l|l|l|l|}
\hline
\begin{tabular}[l]{@{}l@{}}Dataset\\ Size\end{tabular} & \multicolumn{2}{l|}{\begin{tabular}[l]{@{}l@{}}Running Time \\Performance  (minutes)\end{tabular}} & \begin{tabular}[l]{@{}l@{}}Improvement \\ ratio\end{tabular} \\ \cline{2-3}
 & Binary MI     & DDP-MI   &     \\ \hline
3M     & 10.7      & 1.7     & 6.3       \\ \hline
6M     & 13      & 1.8     & 7.2      \\ \hline
30M     & 16         & 2.3        & 7.0  \\ \hline
60M    & 16.4      & 2.4   & 6.8   \\ \hline
\end{tabular}
\caption{Running Time Performance Comparison Between Our Binary and DDP-MI}
\label{tab:performance1}
\end{table}

\begin{figure}[h]
    \includegraphics[width=0.5\textwidth]{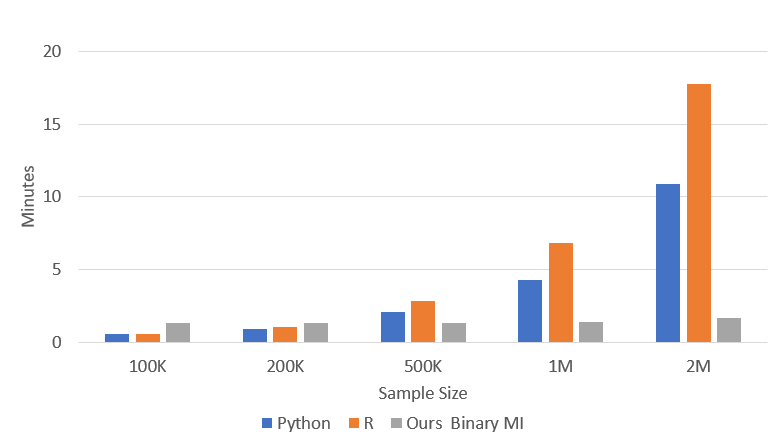}
  \caption{Running Time Performance of Python, R and our Binary MI}
  \label{fig:binaryMI}
\end{figure}

We compare the binary MI performance with the ones in Python and R. Figure~\ref{fig:binaryMI} demonstrates the running time comparison between these three implementations. In this experiment, we start from sample size of 100K to 2M because both Python and R had memory crash for data greater than 2M. Since all implementations are based on binary MI, we only use one partition (years $>$ 2015) in this experiment. It can be seen from this figure, Python's and R's implementations can handle smaller data more efficiently than our MI implementation. However, as data samples increase, their running performance degrades remarkably. 

\begin{figure}[h]
    \includegraphics[width=0.5\textwidth]{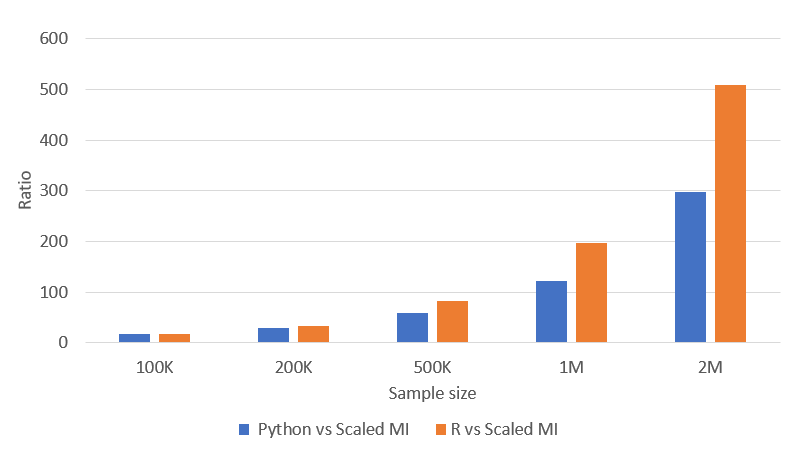}
   \caption{Running Time Performance Improvement of DDP-MI vs Python and R}
   \label{fig:scaledMIRatio}
\end{figure}

We also compared our DDP-MI with both Python and R implementations. 
In this experiment, we used all the years as the partitions (classes). 
For Python and R implementation, we add a loop to run MI on all years using one-against-all technique. 
Figure~\ref{fig:scaledMIRatio} shows the improvement ratio of Python and R compared to our DDP-MI. 
From this figure, it can be concluded that the proposed DDP-MI technique performs very well on very large-scale datasets with many partitions.

To further demonstrate the effectiveness of the DDP-MI on this data, we extracted top features (words) generated by MI for each year. Table~\ref{tab:keywords} lists some of these features for different year ranges. 

\begin{table}[h]
\centering
\begin{tabular}{|l|l|}
\hline
\textbf{Year}   & \textbf{Top MI ranked Keywords} \\ \hline
1970-1980       &  \begin{tabular}[l]{@{}l@{}}     Backtracking, arithmetic, Fourier, graph, assembly,\\ relational db, PASCAL, AI FORTRAN, parser,\\ matrix, grammar             \end{tabular}          \\     \hline
1980-1990       &   \begin{tabular}[l]{@{}l@{}}  Database, trees, algorithms, VLSI, PROLOG,\\ CMOS, silicon, LISP, ATM       \end{tabular}                     \\ \hline
1990-2000       &   \begin{tabular}[l]{@{}l@{}}    Tensor, handwritting, AI, RL, graphG, parallell,\\ Moore, eigenvector, TCI, GA   \end{tabular}                        \\ \hline
2000-2005       &   \begin{tabular}[l]{@{}l@{}}   Windows, genetic, wavelength, optical Internet,\\ filesystem, XML, eGoverment   \end{tabular}                         \\ \hline
2005-2010       &   \begin{tabular}[l]{@{}l@{}}      Fuzzy, ontology, Web, music, speech, languages,\\ RFID,multiagent, XML    \end{tabular}                     \\ \hline
2010-2015           &   \begin{tabular}[l]{@{}l@{}}     Languages, speech, images, social media, SaaS,\\ Robotic, MaPReduce, Kinetic, cloud, agile, sparse,\\ recommendation   \end{tabular}                        \\ \hline
\textgreater{}2015            &   \begin{tabular}[l]{@{}l@{}}  Tracking, encoder, quantum, LSS, navigation, cloud,\\ object, graph, convergence, optimal, regularization  \end{tabular}                          \\ \hline
\end{tabular}
\caption{Distinguishing Keywords using DDP-MI Across Various Years}
\label{tab:keywords}
\end{table}

\section{Conclusions}
We demonstrated a scalable implementation of Mutual Information Ranking that can be used in common distributed computing environments using differentially private SQL. We then applied the implementation in a diverse range of problems and demonstrated sample results. We extended Mutual Information ranking to more than binary partitions and beyond single step.  
To mitigate potential for abuse, this technique should be used only in environments with strong security, policy, and audit controls, in compliance with regulations.  
Common defenses include ``eyes-off'' processing infrastructure, always-encrypted data, and automated expungement. 
On top of these privacy defenses, differential privacy provides defense-in-depth, assuring that released ranking information cannot be used to harm individual privacy.

While the techniques described here are efficient for big data, further improvements in efficiency can be gained through application of probabilistic algorithms.  These algorithms gain efficiency at some cost in accuracy. How these optimizations complement privacy analysis is an interesting area for future research.

\section*{Acknowledgment}

The authors would like to thank Core Data Engineering team in Microsoft Azure COSINE. We work with a lot of partner teams within Microsoft and their collaboration have helped us immensely in developing and applying this methodology on a wide range of business problems. Specifically, we would like to thank Maxime Prat, Marc Mezquita, and Prakhar Panwaria.



\bibliographystyle{IEEEtran}
\bibliography{bare_conf}
%



\end{document}